# Successive spin glass, cluster ferromagnetic, and superparamagnetic transitions in $RuSr_2Y_{1.5}Ce_{0.5}Cu_2O_{10}$ complex magneto-superconductor

Anuj Kumar[1,2], R. P. Tandon[2] and V. P. S. Awana[1,*]

[1]*Quanatum Phenomena and Application Division, National Physical Laboratory (CSIR), Dr. K. S. Krishnan Road, New Delhi-110012, India*

[2]*Department of Physics and Astrophysics, University of Delhi, North Campus New Delhi-110007, India*

We report structural, DC magnetization, detailed linear/non-linear AC susceptibility, (with applied frequency and amplitude) isothermal and thermoremanent magnetization (TRM) behavior for $RuSr_2Y_{1.5}Ce_{0.5}Cu_2O_{10}$ (YRu-1222) magneto-superconductor to understand its complex magnetism. Studied sample is synthesized through the novel solid state High Pressure (6 GPa) High Temperature (1450°C) (HPHT) technique. The compound is crystallized in tetragonal structure with space group *I4/mmm* (No. 139). DC magnetic susceptibility shows that studied YRu-1222 is magneto-superconducting with Ru spins magnetic ordering at around 110 K and superconductivity (SC) in the Cu-$O_2$ planes below ~ 30 K. Frequency and field dependent detailed AC magnetic susceptibility measurements confirms the spin-glass (SG) behavior with homogeneous/non-homogeneous ferromagnetic (FM) clusters in this system. Variation of cusp position with applied AC frequency follows the famous *Vogel-Fulcher law*, which is commonly accepted feature for spin-glass (SG) system with homogeneous/non-homogeneous ferromagnetic clusters embedded in spin-glass (SG) matrix. Above the freezing temperature ($T_f$), first and third harmonics AC susceptibility analysis indicated possibility of the co-existence of spin cluster ferromagnetism with superparamagnetism (SPM). The *M-H* loops at low temperature exhibit the ferromagnetic behavior with rather small coercive field ($H_c$) and remnant magnetization ($M_r$). Summarily, the magnetic (DC and AC) susceptibility measurements and their analysis have enabled us to unearth the complex magnetism in terms of successive SG-FM-SPM transitions with temperature.

**PACS Nos.** 74.70 Pq, 61.66.-f, 75.30Cr, 75.78.-n

**Keywords:** *Ruthenates, Structure, Magnetic Susceptibility, Magnetization dynamics*

# I. Introduction

Superconductivity (SC) and ferromagnetism (FM) are two antagonistic states of matter that tends to avoid each other. The mutual co-existence of these two states below their magnetic ordering temperature ($T_m$) and superconducting transition temperature ($T_c$) are matters of fundamental interest. This antagonistic nature between superconductivity and ferromagnetism has long been recognized. In 1959 it was proposed [1] that co-existence can occur simultaneously, while the ferromagnetic state can adjust itself with a non-uniform structure to accommodate superconductivity. Some non-uniform ferromagnetic states have been observed to co-exist with the superconducting state, which are called ferromagnetic superconductors [2]. These compounds become superconducting below $T_c$ and ferromagnetic at further lower temperature $T_m$, such as $ErRh_4B_4$, $HoMo_6S_8$ and $HoMo_6Se_8$. There are only a few compounds where $T_m$ is higher than $T_c$ and both states do co-exist below superconducting transition temperature ($T_c$). I. Felner and his colleagues [3] reported the co-existence of weak ferromagnetism ($T_m < 100$ K) with superconductivity ($T_c \sim 45$ K) in the high $T_c$ rutheno-cuprates and coined the term 'superconducting ferromagnets' ($T_m > T_c$), in contrast to ferromagnetic superconductors ($T_m < T_c$). The discovery of co-existence of weak ferromagnetism (W-FM) and superconductivity (SC) in $RuSr_2Ln_{1.5}Ce_{0.5}Cu_2O_{10-\delta}$ (Ru-1222) and $RuSr_2LnCu_2O_{8-\delta}$ (Ru-1212) with Ln=Eu, Gd, Sm and Y attracted a lot of attention from scientific community [3-6]. These systems show weak ferromagnetic ordering below the 100-135 K and superconductivity at a lower critical temperature of about 15-40 K, depending on the synthesis and annealing conditions. It seems that the pair breaking phenomenon due to magnetic interactions does not play a role in this system. Unit cell of rutheno-cuprates consists of alternating layers of Ru-$O_2$ and Cu-$O_2$ planes. Ru-$O_2$ layer is responsible for weak ferromagnetism and carrier creation mechanism, while superconductivity resides in Cu-$O_2$ planes below a certain temperature and both of the phenomena are seemingly decoupled from each other. Muon-spin rotation (*ZF-µSR*) for Ru-1222, concluded the possibility of phase separation [7, 8] in terms of various magnetic domains, but supported the idea concerning the co-existence of bulk magnetism and superconductivity in this system. The *ZF-µSR* could detect the different magnetization domains in the bulk material and as result proposed phase separation. Our detailed magnetization results do not go against the phase separation scenario, but rather support the same. We talk of magnetic phase separation here and not the structural phase impurities. Recently, efforts have been devoted to the understanding of phase purity, lattice distortions and true nature of magnetism in Ru-1222 structure [9]. Phase purity, in particular ruling out the presence of $SrRuO_3$ (SRO) magnetic impurity is crucial. In order to support confirmations of microscopic uniform co-existence of superconductivity and magnetism, the presence of impurity phases, in particular the magnetic one need to be rules out. Lattice distortions such as rotations of $RuO_6$ octahedra can also affect the magnetic structure via spin-orbit coupling (Dzyaloshinsky-Moriya interactions) anti-symmetric exchange interactions or single-ion anisotropy [10, 11].

Some unsolved questions still remains about the exact type of magnetic ordering in these systems. Various experimental techniques being used to understand the exact type of magnetic ordering for this system are muon-spin rotation (μ-SR) [7, 8], magnetic resonance (MR) [12], neutron powder diffraction (NPD) [13, 14], magnetization [15, 16] and nuclear magnetic resonance (NMR) [17]. Interestingly, as far as exact magnetic order is concerned, all these techniques are not in full agreement with each other [7, 12-17]. The only commonality was that all the techniques indicated the presence of *canted antiferromagnetic* ordering being embedded in ferromagnetic matrix. Though, some earlier neutron diffraction data [18, 19] did not reveal long range antiferromagnetic (AFM) order, more recent works had clearly exhibited the evidence of the long-range AFM ordering of Ru spins in Ru-1222 phase [20-22]. In fact it is known by now that there is antiferromagnetism (or most likely weak ferromagnetism) in this material down to 2 K [20-22]. Neutron scattering is more authentic tool to determine magnetic structure than the bulk magnetization. Also, it was purposed that in Ru site Nb substituted Ru-1222 compounds there are interacting clusters in the $Ru-O_2$ layers, without any long-range magnetic order [23].

Furthermore, the slow spin dynamics [24] suggested that FM clusters in Ru-1222 could exhibit superparamagnetism (SPM). On the other hand, detailed observations of the frequency and field dependent peak of the *AC* susceptibility as a function of temperature along with isothermal magnetization measurements at low and high fields [25, 26] indicated spin-glass (SG) and cluster ferromagnetism behavior in Ru-1222. As reported for EuRu-1222 system [27], though clear peak shift is seen with frequency in real part of AC susceptibility, the characteristic frequency based upon Arrhenius fit is found to be unrealistic i.e., $10^{130}$ Hz. However, *Vogel-Fulcher* fit suggested SG state with $f_o = 10^{12}$ Hz, which is acceptable. The only question remains is the small deviation of *Vogel-Fulcher* fit at lower side of temperature range [27]. It is well known that frequency dependence of the peak and isothermal magnetization are characteristics features of both spin-glass (SG) and superparamagnetism (SPM), yet the physics behind them is different. Hence, a more careful investigation of the Ru-1222 is required to probe the exact nature of these two states. AC susceptibility and its various harmonics study is very important tool to exactly probe the complex magnetization of such systems. The magnetization measured in presence of excitation AC field can be represents as power law [28, 29],

$$M = M_0 + \chi_1 H^1 + \chi_2 H^2 + \chi_3 H^3 + \chi_4 H^4 + \ldots \quad (1)$$

Where $\chi_1$, $\chi_2$, $\chi_3$ are the first, second, third order harmonics of AC susceptibility respectively, which provides useful information about the existing complex magnetism of the studied system.

Superparamagnetism (SPM) is a magnetic system consisting of widely spaced (isolated) or non-interacting single domain particles (macro-spins or super-spins) having a magnetic moment $m \sim 10^3$-$10^5$ $\mu_B$, which act as independent isolated particles. Although in an ensemble of isolated magnetic particles, direct quantum exchange interactions between them may be negligible, the magnetic properties are determined by the dipole field energy along with the thermal and magnetic anisotropy energies [30, 31]. These isolated domains are separated by domain walls to minimize the net free energy of the system. The magneto-static energy increases proportionally to the volume of the material, while the domain wall energy increases proportionally to the surface area. Though, magnetic ordering can still exist within the clusters, resulting in large saturation magnetization ($M_{sat}$), the random orientation of the clusters could reduce the coercivity ($H_c$) and remanent magnetization ($M_{rem}$) to zero [32]. In the superparamagnetic state, below a certain temperature, called the blocking temperature ($T_B$), the anisotropy energy crosses the thermal energy, so that easy axis of magnetization in clusters is oriented in random direction.

The part of present YRu-1222 was reported earlier but the AC susceptibility measurements were done only in narrow range of frequency (165-1465 Hz) [33], which could not detect the peak shift with frequency and hence excluded the SG state. In present work the AC susceptibility studies are carried out from 33 Hz to 9999 Hz and clear peak shift is seen, which could be fitted with Vogel-Fulcher law representing SG state. We believe the extended frequency range AC susceptibility measurements were warranted on YRu-1222. Hence in the present work, the temperature dependence of the DC magnetization, linear and non-linear AC susceptibility with first and higher harmonics and isothermal magnetization at high magnetic field of $RuSr_2Y_{1.5}Ce_{0.5}Cu_2O_{10}$ (YRu-1222) sample are investigated in detail to understand the spin-glass (SG), ferromagnetic clusters and superparamagnetism (SPM) behavior in the system. First, some DC magnetization and linear AC susceptibility with frequency is performed, which confirmed the spin-glass (SG) state. Second, it will be shown that YRu-1222 is better described as a spin-glass (SG) with magnetic clusters as a result of the formation of homogeneous/non-homogeneous ferromagnetic clusters in spin-glass (SG) matrix. Third, fitting of first and third harmonics of AC susceptibility with Wohlfarth's model (WM) suggests the superparamagnetism (SPM) state in YRu-1222. All results are discussed in a systematic way. Summarily, complex magnetism of superconducting ferromagnet YRu-1222 is unearthed. Note that the lower case subscript notation is for the superconducting transition temperature ($T_c$), and the upper case subscript notation for Curie temperature ($T_C$), whereas the magnetic ordering transition is marked by $T_m$.

## II. Experimental Details

Polycrystalline sample of chemical composition $RuSr_2Y_{1.5}Ce_{0.5}Cu_2O_{10}$ was synthesized through standard High Pressure High Temperature (HPHT) solid state reaction route under optimized 6 GPa pressure and 1450°C temperature. For the HPHT synthesis the ratio of the ingredients used are $(RuO_2) + 2(SrO_2) + 3/4(Y_2O_3) + 1/2(CeO_2) + 3/4(Cu_2O) + 1/2(Cu)$ resulting in $RuSr_2Y_{1.5}Ce_{0.5}Cu_2O_{10}$ (YRu-1222). The ingredients were mixed in an agate mortar with pestle in a Glove Box to obtain starting material for high pressure synthesis. Later on, around 300 mg of the raw mixture was sealed in a high purity gold capsule and allow to heat in a flat belt type HPHT apparatus at 6 GPa and 1450°C for 3 h. After the heat treatment, the sample was quenched to room temperature, and the pressure was slowly released [34]. To confirm the exact oxygen content in the synthesized sample, the weight of the capsule was measured before and after the high pressure reaction. No reasonable change was observed in the weight, warranting the fixed nominal oxygen content of 10.0 in the synthesized sample. The surface of the sintered high-pressure sample was polished with sand paper; the inner clean black sintered material was used for characterization.

The structure and phase purity of HPHT synthesized sample was confirmed by X-ray diffraction (XRD) measured at room temperature in the scattering angular ($2\theta$) range of 20°-80° in equal steps of 0.02° using Rigaku Diffrectrometer with Cu $K_\alpha$ ($\lambda = 1.54$ Å) radiation. Detailed rietveld analysis was performed using the *FullProf* program. Detailed DC and AC (linear and non-linear) susceptibility data were measured on physical property measurements system (PPMS-14T, Quantum Design-USA) in temperature range 1.9 – 200 K. The isothermal magnetization (*M-H*) loops at different temperatures with applied magnetic field up to ± 5 kOe were also measured using the same PPMS. Detailed linear and non-linear AC susceptibility as a function of temperature T, in (i) in the frequency ranges of 33-9999 Hz and (ii) in the AC drive field amplitude 1-17 Oe in the zero external DC magnetic fields, were also measured on physical property measurements system (PPMS-14T, Quantum Design-USA).

## III. Results and Discussion

The phase purity of rutheno-cuprates sample is very important because a minute impurity of magnetic phase $SrRuO_3$ (SRO) and $Sr_2YRuO_6$ ($2110_6$), which readily tends to form in rutheno-cuprates matrix, can alter the net outcome magnetization. After several optimizations through HPHT method we obtained the phase pure YRu-1222 compound [34]. The sample is HPHT synthesized and despite various trials minute amount of impurity of double perovskite ($2110_6$) phase is seen. Figure 1 depicts the room temperature observed and calculated X-ray diffraction (XRD) patterns of studied $RuSr_2Y_{1.5}Ce_{0.5}Cu_2O_{10}$ (YRu-1222). The structural analysis was performed using the Rietveld refinements with help of *FullProf* software. Rietveld analysis confirmed the single phase formation of studied YRu-1222 compound in space group *I4/mmm*. All Rietveld refined parameters (Lattice parameters, Wyckoff position and site occupancy) of studied YRu-1222 compound are shown in Table I.

Figure 2 depicts the DC magnetization (*M-T*) of HPHT synthesized $RuSr_2Y_{1.5}Ce_{0.5}Cu_2O_{10}$ (YRu-1222) compound in zero-field-cooled (ZFC) and field-cooled (FC) situations measured at 20 Oe. The compound exhibits complex magnetic behavior. We define the $T_C$ (Curie temperature), simply a paramagnetic (PM) to ferromagnetic (FM) transition, as the temperature corresponding to common tangent on ZFC and FC curve cutting the temperature *x*-axis. The $T_C$ (Curie temperature) of YRu-1222 is around 110 K as marked in figure 2. In fact the branching of FC and ZFC starts with a dip before $T_C$ (110 K) at around 120 K ($T_N$). This is consistent with an earlier report on YRu-1222 where it is shown that before canted ferromagnetism the Ru spins order antiferromagnetically (AFM) [35]. Interestingly the AFM persists down to 2 K as evidenced from NPD [20-22]. Below the $T_C$ (Curie temperature) at around 90 K, the ZFC and FC curves branch out. This temperature, where ZFC and FC curves branch out called the freezing temperature below which the system enters into a new state called glassy state. Below the freezing temperature ($T_f$) the magnetic moment corresponding to ZFC curve decreases and FC curve increases with decreasing the temperature. The system enters into a superconducting state below the superconducting transition temperature $T_c$ = 28 K (a kink observed in both ZFC and FC curves shown in figure 2) and finally a diamagnetic transition at $T_d$ = 21 K. Despite a clear diamagnetic transition in ZFC at $T_c$, the FC branch shows a paramagnetic Meissner effect (PME) like situation. The PME is often seen in some superconductors [36]. For more detailed description, please see Refs. [25, 37-38]. The strong irreversibility between the ZFC and FC curve exhibits in the *M-T*, typical of a superparamagnetism (SPM) relaxation phenomenon of a spin glass/cluster glass system [39-41]. Several features, which are shown here, support the likely occurrence of a spin glass/cluster glass in $RuSr_2Eu_{1.5}Ce_{0.5}Cu_2O_{10-\delta}$ (EuRu-1222) sample, and are similar to the results for some other compositions of this system as being reported by other authors [42]. The freezing temperature ($T_f$) is also observed in AC susceptibility measurements, which will be discussed in later sections. M-H loops for studied YRu-1222 at 5 K and 20 K are shown for an applied field range (-3000 Oe ≤ H ≤ +3000 Oe) in the inset of figure 2. The compound exhibits clear ferromagnetic magnetization loops below the Curie and superconducting transition temperatures with a reasonable coercive field ($H_c$) and remnant magnetization ($M_r$). The value of $M_r$ and $H_c$ decreases monotonically as the temperature increases.

Figure 3 shows the typical isothermal magnetization of HPHT synthesized YRu-1222 compound at various temperatures (5, 20, 50, 75, 100, 125, 150 and 200 K) for applied magnetic field range (-50 kOe ≤ +50 kOe). At low temperatures the *M-H* loop exhibits the *S* type shape with reasonable coercive field ($H_c$) and remnant magnetization ($M_r$), which are the characteristics features of the spin-glass (SG) system. The opening of *M-H* loop at low temperature resembles the ferromagnetic nature in the system. Seemingly the system shows the co-existence of spin-glass (SG) and ferromagnetism. The magnetization becomes a non-linear function of applied field and shows ferromagnetic behavior with hysteresis loop at low field range. The isothermal magnetization as a function of applied field at 5 K may be viewed as: $M(H) = \chi H + \sigma_s(H)$, where $\chi H$ is the linear contribution from the antiferromagnetic ($T_C$) Ru spins and $\sigma_s(H)$ represents the ferromagnetic component of Ru. The appearance of ferromagnetic component at low temperature within antiferromagnetic/spin glass Ru spins is possibly due to the slight canting of Ru spins as seen from neutron diffraction for Ru-1212 [19]. As the temperature increases this S type shape of *M-H* loop transforms to the linear paramagnetic (*PM*) shape. At 200 K the *M-H* is like a straight line, resembling the PM nature of compound at that temperature. It is also observed that these *M-H* loops do not saturate even at 50 kOe applied magnetic field. Both, the absence of magnetization saturation at high field and the existence of hysteresis loop at low temperatures and low field-regions, are the characteristics of spin-glass (SG) [43, 44] phase with possibly co-existing ferromagnetic clusters. This co-existence of spin-glass (SG) and ferromagnetic clusters will be discussed in later sections in details. A complementary test is shown in figure 4 where an Arrott plot, [45] $M^2$ vs. *H/M* is performed for the same set of isothermal magnetization curves. In this standard experimental method the occurrence of FM order is predicted to occur when straight line $M^2$ α *H/M* are obtained in the plots. Further, it defines the Curie temperature ($T_C$) of the isotherm whose linear extrapolation intercepts the vertical axis at zero. In our case we find $T_C$ ~ 100 K for the studied YRu-1222 sample, which is in agreement with the $T_C$ = 110 K observed from the *M-T* curve (figure 2). No sign of spontaneous magnetization are observed in Arrott plots instead of that there is strong curvature towards *H/M* axis and some intercept on $M^2$ axis. Absence of spontaneous magnetization confirms the short-range magnetic ordering [46], which is also a feature of spin-glass (SG) with homogeneous/non-homogeneous ferromagnetic clusters.

A further investigation of spin-glass/cluster-glass behavior is also done by thermo-remnant magnetization (TRM) measurements. The time response of DC magnetization is important to reveal the spin dynamics for spin-glass (SG) system [28, 29]. The behavior of a spin-glass (SG) below $T_f$ is irreversible and complicated by the aging process, so it is necessary to employ a well-defined *H-T* procedure to obtain a meaningful data. The sample was field-cooled (FC) in the presence of 5000 Oe field from 200 K to 60 K and after certain waiting times ($t_w$ = 100 s and 500 s), the field was reduced to zero and the corresponding decay of magnetization was recorded as a function of elapsed time. The result for the studied YRu-1222 is shown in figure 5. The observed behavior of TRM is strictly the same as those of site-disordered spin-glasses system [47]. Longer the hold time $t_w$, the slower the decay of the TRM. The system has become "stiffer" with time. The changes observed in *M (t)* measured for different values $t_w$ shows the occurrence of aging effects, which means that the system is in meta-stable spin-glass (SG) state. The situation will be clearer in next sections, when we will discuss the *AC* susceptibility and its higher harmonics to explore the possible presence of spin-glass/cluster ferromagnetism/superparamagnetism phenomena in studied YRu-1222.

The AC susceptibility (linear and non-linear) technique is a powerful method, which has been used to study the spin-glass/cluster spin-glass/superparamagnetism type systems. Both real and imaginary parts exhibit sharp frequency dependent cusp according to the desired phenomena (spin-glass/cluster spin-glass/superparamegnetism). It is well known that a small external DC magnetic field as low as few mili-Orested (m-Oe) can change the cusp nature. The main panels of figure 6(a) and 6(b) show the temperature dependence of real part $\chi'$ (dispersion) and the imaginary part $\chi''$ (absorption) of the first harmonics of $AC$ susceptibility $\chi_{ac}$, in the presence of applied frequency (33, 333, 666, 999, 3333, 6666 and 9999 Hz) at zero external DC field. Before major magnetic transitions (SG/FM peak in magnetization), the AFM correlations related Neel temperature $T_N$ is seen clearly in both Figures 6(a) and 6(b) at around 120 K, again consistent with an earlier report on this system [35]. Further as mentioned earlier, this AFM order is reported persistent down to 2 K as evidenced from NPD results [20-22]. Neel temperature ($T_N$) does not shift with the frequency of applied AC field. Inset of figure 6(a) and 6(b) show the enlarge view of real $\chi'$ and imaginary $\chi''$ part of AC susceptibility respectively. Real ($\chi'$) and imaginary ($\chi''$) both part show the clear peak around the spin-glass (SG) transition or peak temperature ($T_p$) or freezing temperature ($T_f$). $T_p$ is an average blocking temperature where the clusters moments begin to freeze, and $T_f$ is the freezing temperature where this thermally activated process reaches a maximum. Possibly the clusters consists of FM or FM-like islands in an AFM matrix [48, 49]. This would be consistent with the fact that $T_p$ and $T_f$ are always smaller than $T_C$ for studied system. This peak temperature corresponds to the peak in the ZFC curve with a slight change in peak temperature because of the difference in the response of the system to DC and AC fields. Inset of figure 6(a) shows that height of the peak corresponding to the freezing temperature ($T_f$) decreases and also the peak shifted towards higher temperature with increasing the frequency ($f$). Similarly, for imaginary part ($\chi''$) the height of the peak decreases and shifted towards the higher temperature (see inset of figure 6(b)). However, the qualitative effect is same but the exact shift is larger for imaginary part ($\chi''$) than the real one ($\chi'$). It is observed there is a change in freezing temperature ($T_f$) with applied frequency. The change in freezing temperature $T_f$ ($\chi'$) ($T_f$ = 89.8 K at $f$ = 33 Hz and $T_f$ = 90.7 K at $f$ = 9999 Hz) with applied frequency is the characteristics of spin-glass (SG) behavior. At primary stage it is estimated from the quantity $k$ =

$\Delta T_f/T_f \Delta (\log_{10} f)$, where $\Delta$ represents the change in the corresponding quantity. It is known this quantity ($k$) varies in the range of 0.004-0.018 for spin-glass (SG) system, however for super-paramagnetic systems it is of the order of 0.3-0.5 [27]. $T_f$ is as assumed the temperature corresponding to the maximum value of the $\chi'$ curve or the inflection point from the $\chi''$ curves. Here we obtained $k = 6.6 \times 10^{-3}$ or 0.0066, which is in good agreement with the typical spin-glass (SG) system values, e.g., $2 \times 10^{-2}$, $1.8 \times 10^{-2}$ and $6 \times 10^{-3}$ for $La(Fe_{1-x}Mn_x)_{11.4}Si_{1.6}$, NiMn and $(Eu_{1-x}Sr_x)S$ respectively [50, 51]. Hence, it is clear that the studied YRu-1222 system is a typical spin-glass (SG) with homogenous/non-homogenous ferromagnetic clusters just below the freezing temperature ($T_f$), to be discussed in later sections. There are basically two different possible interpretations of the spin-glass freezing: first one is the existence of true equilibrium phase transition at a fixed temperature (canonical spin-glass) [52] and the second assumes the existence of ferromagnetic homogeneous/non-homogeneous clusters embedded in AFM matrix with non-equilibrium freezing [53]. To further verify the spin-glass (SG) state or magnetically interacting clusters state in the studied YRu-1222, the *Vogel-Fulcher law* [28, 29] purposed,

$$\omega = \omega_o \, exp. \, [-E_a/k_B(T_f - T_o)] \ldots (2)$$

where, $E_a$ is the activation energy or the potential barrier separating two nearby clusters, $\omega_o$ is the characteristics individual frequency of clusters, $T_f$ the freezing temperature and $T_o$ is the *Vogel-Fulcher* temperature, which gives inter-clusters interaction strength. When $T_o = 0$ means there is no inter-clusters interaction (isolated clusters or superparamagnetism state) takes place then the *Vogel-Fulcher law* transforms into the well known *Arrhenius law* [28, 29], which is useful to determine the relaxation process of non-interacting magnetic clusters,

$$\omega = \omega_o \, exp. \, [-E_a/k_B T_f] \ldots (3)$$

The *Vogel-Fulcher law* fits with experimental data of studied YRu-1222 for various characteristics frequency ($\omega_o/2\pi$) ranging $10^{10}$-$10^{13}$ Hz (shown figure 7). Figure 7 depicts a linear fit between the freezing temperature $T_f$ and $1/(lnf_o/f)$. Two parameters (Activation energy $E_a$ and *Vogel-Fulcher* temperature $T_o$) are calculated for corresponding to each chosen characteristic frequency. The value of the inverse of slope $E_a/K_B$, *Vogel-Fulcher* temperature $T_o$ and the parameter $t^* = (T_f - T_o)/T_f$, corresponding to each characteristic frequency ranging from $10^{10}$-$10^{13}$ Hz are listed in Table II. The values of the *Vogel-Fulcher* temperature $T_o = 88.34$ K, 87.93 K,

87.82 K and 87.52 K corresponding to the each characteristic frequency $10^{10}$ Hz, $10^{11}$ Hz, $10^{12}$ Hz and $10^{13}$ Hz respectively, are in good agreement with the value of freezing temperature $T_f$ = 90.70 K, obtained from the AC susceptibility measurements. Also, for spin- glass (SG) system, the parameter t* = $(T_f - T_o)/T_f$ must be < 0.10 and t* ≥ 0.5 for a cluster spin-glass (CSG) system [27, 36]. In our case t* is of the order of 0.026-0.035, qualifying for a pure spin-glass (SG) system. Hence the fitted experimental data of Vogel-Fulcher law and parameter t* indicate the presence of spin-glass (SG) state in the studied YRu-1222 system.

To further verify the spin-glass (SG) state with ferromagnetic clusters, we performed the linear AC susceptibility measurements on studied YRu-1222 system with varying AC drive field and a fixed frequency of 333 Hz. Figure 8(a) and 8(b) revels the real ($\chi'_1$) and imaginary part ($\chi''_1$) part of the first harmonic of AC susceptibility respectively. Both real ($\chi'_1$) and imaginary part ($\chi''_1$) are measured as a function of temperature from range 200 K to 2 K range with zero external DC field bias. Both parts (real and imaginary) are measured with varying AC drive amplitude from 1 Oe to 17 Oe and at a fix frequency 333 Hz. It is observed from figure 8(a) and 8(b) that the peak temperature corresponding to $\chi'_1$ and $\chi''_1$ shifts slightly towards the lower temperature and the height of the peak increases with increasing the AC drive amplitude of field. This is a contradictory behavior, because it was observed and well known earlier that the height of the peak decreases with increasing amplitude of the AC drive field for a typical spin-glass (SG) system [28, 29]. When the amplitude of the AC drive field increases, the magnetic energy associated with external AC field is large enough to compare with the thermodynamic energy of the magnetic dipole inside the system. In present situation (figure 8(a) and 8(b)), because there is no freezing of magnetic moments taking place in the direction of applied magnetic field, hence the studied system YRu-1222 is not a pure spin-glass (SG) system. Instead of it has pure spin-glass (SG) state along with some non-interacting homogeneous/non-homogeneous magnetic clusters component. On the basis of above discussed results there may be a possibility of superparamagnetism state in the studied YRu-1222 system, which will be discussed in details in next section.

To confirm the existence of superparamagnetism (SPM) state in the studied YRu-1222 system, we analyzed our AC susceptibility data by using Wohlfarth's model [54]. According to the Wohlfarth's superparamagnetic blocking model, real part of the first harmonic of AC susceptibility ($\chi'_1$) above the blocking temperature ($T_B$) should follow the Curie-Weiss law, while it is independent below the $T_B$. Similarly, real part of the third harmonic of AC susceptibility ($\chi'_3$) above the $T_B$ follows the negative $1/T^3$ dependence and again it is independent below the $T_B$.

$$\chi'_1 = \frac{\varepsilon M_{sat}^2 V}{3k_B T} = \frac{P_1}{T} \; or \; \chi'_1 \propto \frac{1}{T} \ldots \qquad (3)$$

$$\chi'_3 = -\frac{\varepsilon M_{sat}}{45}\left(\frac{M_{sat}V}{k_B T}\right)^3 = -\frac{P_3}{T^3} \; or \; \chi'_3 \propto \frac{1}{T^3} \ldots \qquad (4)$$

Here $\varepsilon$ is the volume fraction occupied by the magnetic particles, $T$ is absolute temperature, $M_{sat}$ is saturation magnetization, $k_B$ is Boltzmann constant and $V$ is the volume of magnetic particles. $P_1$ and $P_3$ are two temperature-independent constants. Figure 9(a) and 9(b) depicts the first order harmonic ($\chi'_1$) and third-order harmonic ($\chi'_3$) of AC susceptibilities as s function of $T^{-1}$ and $T^{-3}$ respectively. Also the solid lines are best fit of equation (3) and (4) to be the experimental data respectively. The $T^{-3}$ temperature dependence of third harmonic ($\chi'_3$) can be reasonably fit with the temperature interval between 88.5 K and 102 K. A similar fit has been obtained over narrower temperature range around 10 K in a study of SPM in polycrystalline $Li_{0.5}Ni_{0.5}O$ compound [55]. On the basis of above mentioned fitting analysis, the range of the fitting the equations is typically around 10 K above $T_B$. Above this temperature the spin correlation within a particle vanishes. Generally, for a conventional SPM system the spin freezing temperature or particle's spin correlation temperature is much higher than $T_B$. For example in superparamagnetic clusters of magnetite $Fe_3O_4$, the Curie temperature of bulk magnetite is around 850 K while the blocking temperature is observed around 20 K [56]. Hence in principle, one can measure the superparamagnetic (SPM) state above $T_B$, covering a large temperature range. But in our case, the blocking temperature and spin freezing temperature are very near to each other, and hence fitting is done in limited range (88.5 K to 102 K). Figure 10 depicts a plot between $1/\chi_{ac}$ vs. T for studied YRu-1222 system, which clearly shows two distinct slopes. Hence it is clear that a superparamagnetic state is developed over a narrow temperature range $T_C \geq T \geq T_f$. It concludes that superparamagnetic (SPM) state co-exists with the spin-glass (SG) state in studied YRu-1222 system.

The value of $H_C$ and $M_{rem}$ are almost zero within the temperature range $T_C \geq T \geq T_f$ but the $M_{sat}$ has definite value (0.30 $\mu_B$) as shown in figure 11. The reason is that just below $T_C$ the FM order lying within the clusters, leading to a non-zero value of $M_{sat}$. On the other hand magnetocrystaline energy become of the order of thermal activation energy of the clusters, resulting the random orientation of the clusters with respect to each other. This random orientation of clusters reduces $H_c$ and $M_{rem}$ to zero. A same feature is observed for superparamagnetic (SPM) particles [29, 57]. Fitting of Wohlfarth's model (WM) of SPM to the AC susceptibility data of YRu-1222 in the range of $T_C \geq T \geq T_f$ suggests that in a particular range of temperature, spin-glass (SG) state with ferromagnetic clusters giving rise to the SPM state. Thus YRu-1222 has different FM phases below the $T_C$.

## IV. Conclusions

In this paper we have reported detailed results and analysis of structural, DC/linear and non-linear AC magnetization, isothermal magnetization and thermoremenet magnetization (TRM) on HPHT synthesized $RuSr_2Y_{1.5}Ce_{0.5}Cu_2O_{10}$ (YRu-1222) magneto-superconductor. The YRu-1222 has a rich verity of magnetic phenomena. A paramagnetic(PM) to antiferromagnetic (AFM) transition at 120 K, canted ferromagnetic (FM) transition at around 110 K, spin-glass (SG) transition temperature ($T_f$) at around 88.5 K, formation of homogeneous/non-homogeneous ferromagnetic non-interacting clusters just below the spin-glass (SG) temperature and also the possible presence of superparamagnetic state in the compound (SPM). The DC and AC susceptibility studied presented in this paper shows that SG state co-exist with some homogeneous/non-homogeneous ferromagnetic non-interacting clusters followed by possible SPM state. The cluster ferromagnetism could be originated from the canting of reported long-range AFM order in this system. The temperature variation of first and third-order harmonic AC susceptibility is fitting well to Wohlfarth's model (WM) of superparamagnetism in a narrow temperature range. A FM cluster and spin-glass (SG) state have been seen to co-exist just below the $T_f$. The superparamagnetism (SPM) and spin-glass (SG) state co-exist between the temperature range $T_C \geq T \geq T_f$. Superconductivity is not affected by various co-existing magnetic phenomena. In last, our results support the presence of spin-glass (SG) state with non-homogeneous ferromagnetic clusters followed by SPM state in YRu-1222 system. Possible random distribution of $Ru^{5+}$-$Ru^{5+}$, $Ru^{4+}$-$Ru^{5+}$ and $Ru^{4+}$-$Ru^{4+}$ exchange interactions may be responsible for observed spin-glass (SG) with ferromagnetic clusters (FM) followed by superparamagnetism (SPM) complex magnetic state.

## Acknowledgements

The authors from NPL would like to thank Prof. R. C. Budhani (Director, NPL) for his keen interest in the present work. One of us, Anuj Kumar, would also thank Council of Scientific and Industrial Research (*CSIR*) New Delhi, Government of India, for financial support through Senior Research Fellowship (*SRF*). Dr. V. P. S. Awana would like to thank Prof. E. Takayama-Muromachi, NIMS Japan for using the HPHT facility. This work is also partially financial supported by *DST-SERC* (Department of Science and Technology) New Delhi, Government of India, funded project on "Investigation of pure and substituted Rutheno-cuprate magneto-superconductors in bulk and thin film form at low temperature and high magnetic field".

# Figure Captions

**Figure 1** Observed (*solids circles*) and calculated (*solid lines*) XRD patterns of $RuSr_2Y_{1.5}Ce_{0.5}Cu_2O_{10}$ compound at room temperature. *Solid lines* at the bottom are the difference between the observed and calculated patterns. *Vertical lines* at the bottom show the position of allowed Bragg peaks.

**Figure 2** *ZFC* and *FC DC* magnetization plots for $RuSr_2Y_{1.5}Ce_{0.5}Cu_2O_{10}$, measured in the applied magnetic field, $H$ = 20 Oe. Inset shows the $M$ vs. $H$ plot at temperature 5 and 20K in the range of -3000 Oe ≤ H ≤ +3000 Oe.

**Figure 3** Typical magnetization loops as a function of applied magnetic field measured at different temperatures (5, 20, 50, 75, 100, 125, 150 and 200 K) in the range - 50 kOe to + 50 kOe.

**Figure 4** Arrott plots ($H/M$ vs. $M^2$) using *DC* magnetization vs. applied field data observed at different fixed temperatures (5, 20, 50, 75, 100 and 125 K).

**Figure 5** Thermoremanent magnetization (TRM) relaxation for T = 60 K and for waiting time $t_w$ = 100 s and 500 s.

**Figure 6(a)** Temperature dependence of the real part of *AC* susceptibility, measured at different frequency with zero external *DC* magnetic fields. Inset shows the enlarged view of the real part of the first harmonic *AC* susceptibility.

**Figure 6(b)** Temperature dependence of the imaginary part of *AC* susceptibility, measured at different frequency with zero external *DC* magnetic fields. Inset shows the enlarged view of the imaginary part of the first harmonic *AC* susceptibility.

**Figure 7** The variation of the freezing temperature $T_f$ with the frequency of the *AC* field, at different characteristics frequency, in a *Vogel-Fulcher* plot. The solid lines are the best fit of equation.

**Figure 8(a)** Temperature dependence of the real part of *AC* susceptibility measured at different amplitude with zero external *DC* magnetic fields.

**Figure 8(b)** Temperature dependence of the imaginary part of *AC* susceptibility, measured at different amplitude with zero external *DC* magnetic fields.

**Figure 9(a)** First order harmonics of *AC* susceptibility is fitted to Wohlfarth's model above the freezing temperature ($T_f$) for studied YRu-1222. The solid red line shows $T^{-1}$ fit to $\chi_1$.

**Figure 9(b)** Third order harmonics of *AC* susceptibility is fitted to Wohlfarth's model above the freezing temperature ($T_f$) for studied YRu-1222. The solid red line shows $T^{-3}$ fit to $\chi_3$.

**Figure 10** The inverse of first-order *AC* susceptibility ($\chi$) plotted with temperature indicating two distinct slopes corresponding to paramagnetic and superparamagnetic phase.

**Figure 11** Saturation magnetization ($M_{sat}$), remanent magnetization ($M_{rem}$) and coercive field ($H_c$) as a function of temperature.

**Table I.** Atomic coordinates and site occupancy of $RuSr_2Y_{1.5}Ce_{0.5}Cu_2O_{10}$

Space group: *I4/mmm*, Lattice parameters; $a$ = 3.8181 (3) Å, $c$ = 28.4951 (7) Å, $\chi^2$ = 2.22

| Atom | Site | $x$ (Å) | $y$ (Å) | $z$ (Å) |
|---|---|---|---|---|
| Ru | 2b | 0.0000 | 0.0000 | 0.0000 |
| Sr | 2h | 0.0000 | 0.0000 | 0.4211 (5) |
| Y/Ce | 1c | 0.0000 | 0.0000 | 0.2932 (8) |
| Cu | 4e | 0.0000 | 0.0000 | 0.1451 (3) |
| O(1) | 8j | 0.8228 (3) | 0.5000 | 0.0000 |
| O(2) | 4e | 0.0000 | 0.0000 | 0.0719 (5) |
| O(3) | 8g | 0.0000 | 0.5000 | 0.1460 (6) |
| O(4) | 4d | 0.0000 | 0.5000 | 0.2500 |

**Table II.** Inverse of slope $E_a/k_B$, Vogel-Fulcher temperature $T_o$ and parameter $t^* = (T_f-T_o)/T_f$

| Characteristic frequency | Inverse of slope $E_a/k_B$ (K) | Vogel-Fulcher temperature $T_o$ (K) | Parameter $t^* = (T_f-T_o)T_f$ |
|---|---|---|---|
| $f_o = 10^{10}$ Hz | 34.60 K | 88.34 K | 0.026 |
| $f_o = 10^{11}$ Hz | 48.31 K | 87.93 K | 0.031 |
| $f_o = 10^{12}$ Hz | 57.80 K | 87.82 K | 0.032 |
| $f_o = 10^{13}$ Hz | 68.97 K | 87.52 K | 0.035 |

**Figure 1**

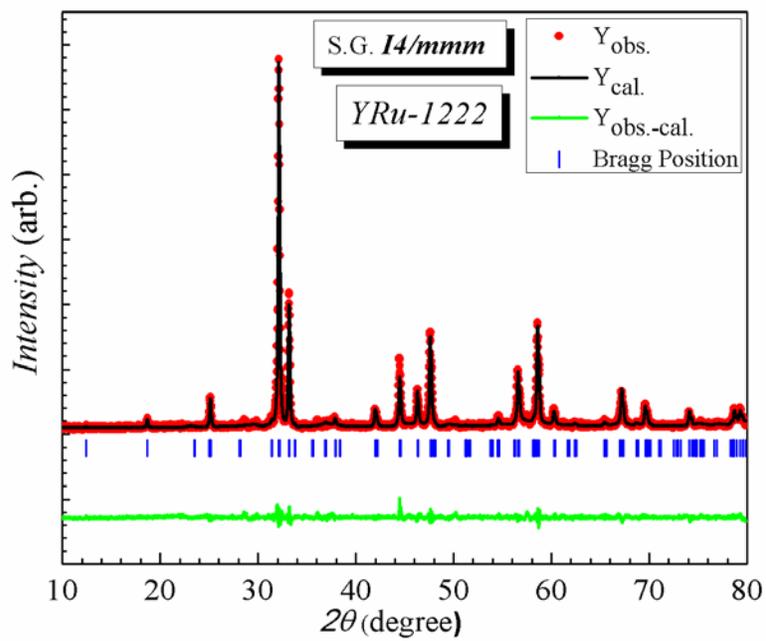

**Figure 2**

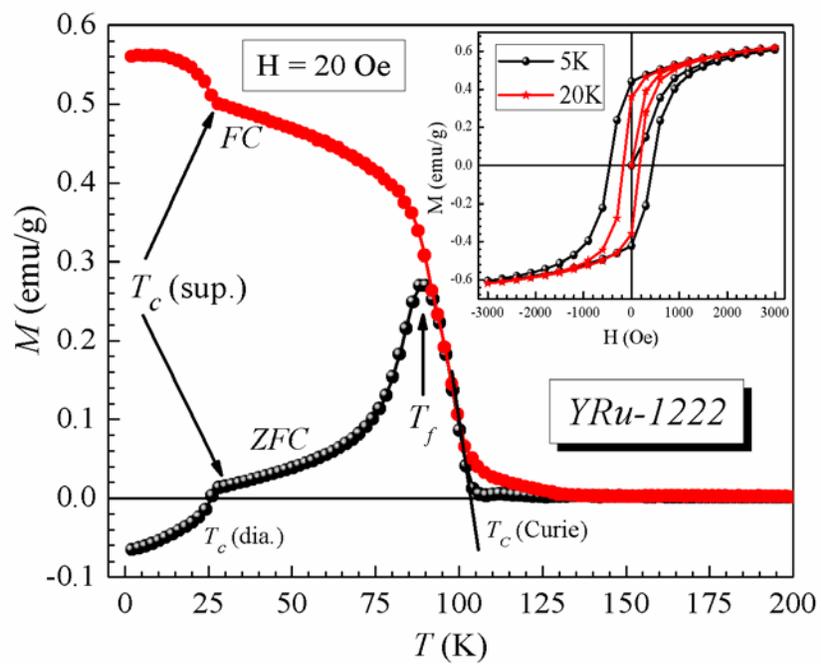

**Figure 3**

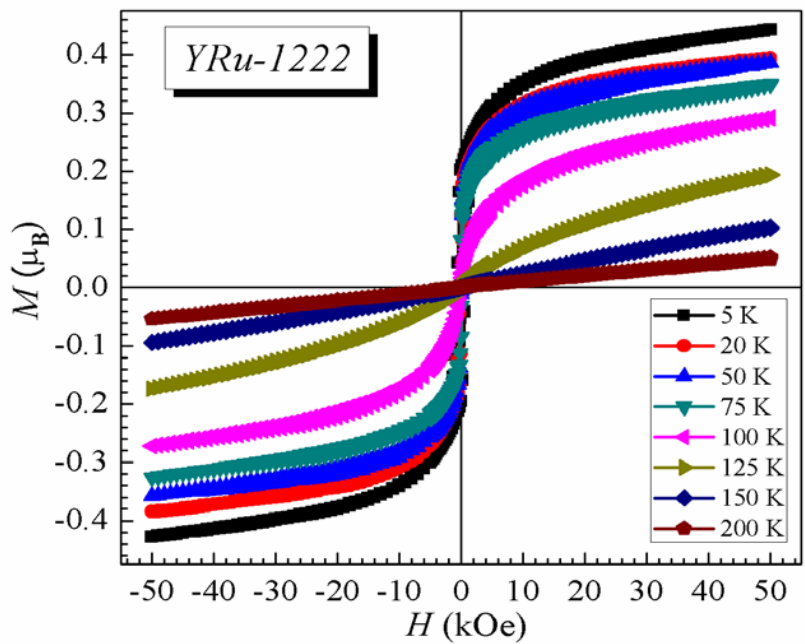

**Figure 4**

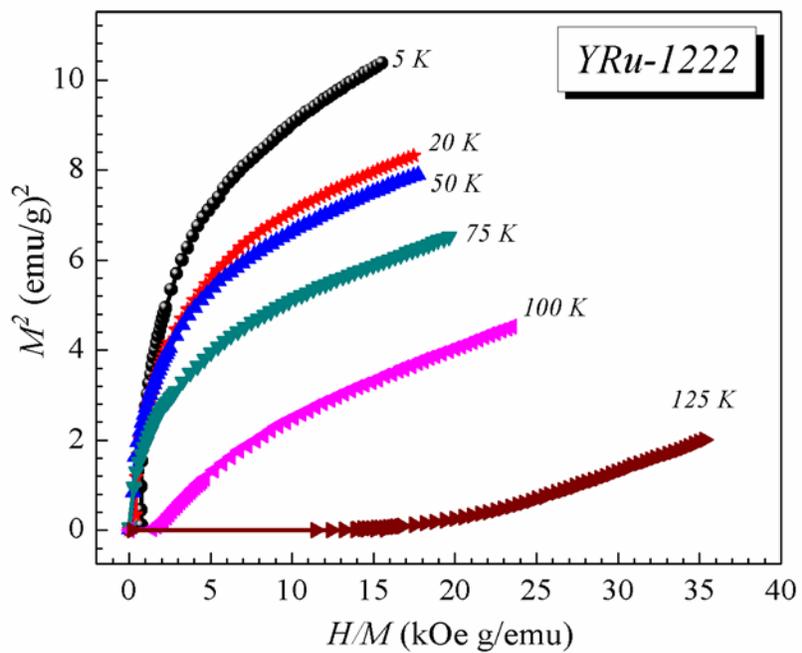

**Figure 5**

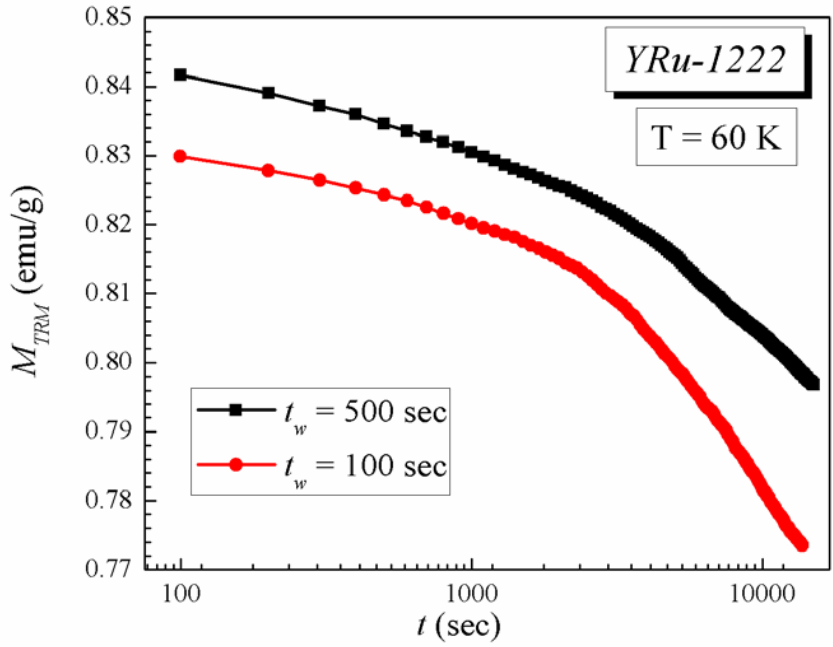

**Figure 6(a)**

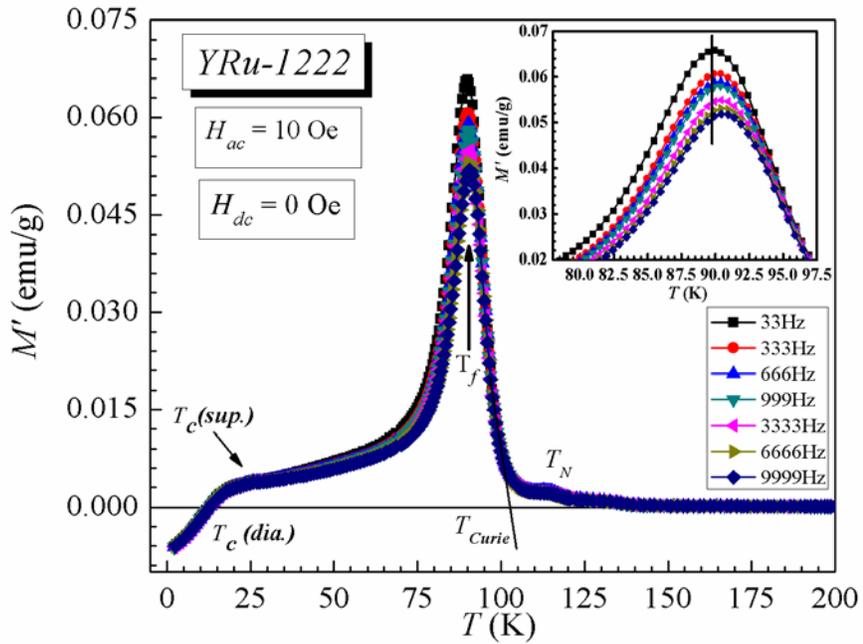

**Figure 6(b)**

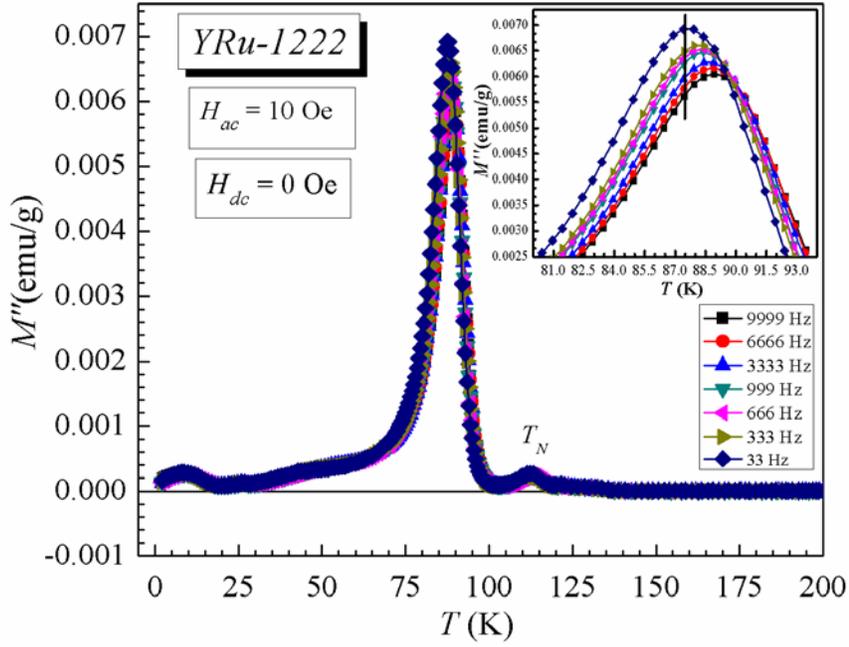

**Figure 7**

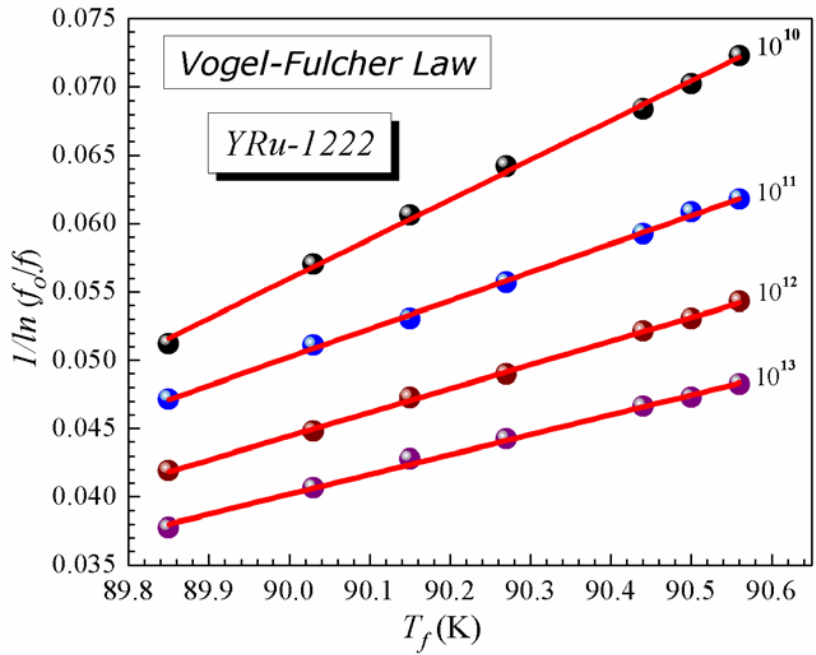

**Figure 8(a)**

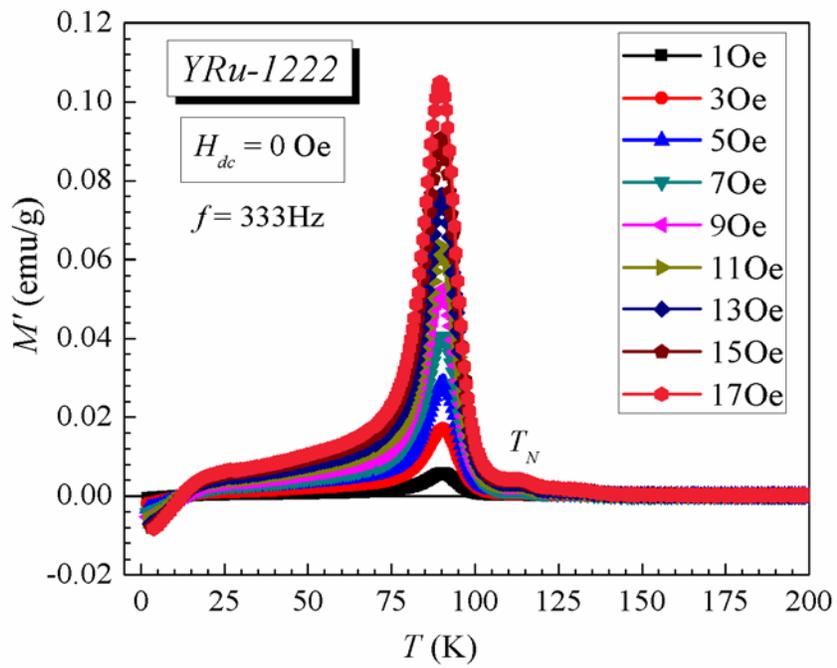

**Figure 8(b)**

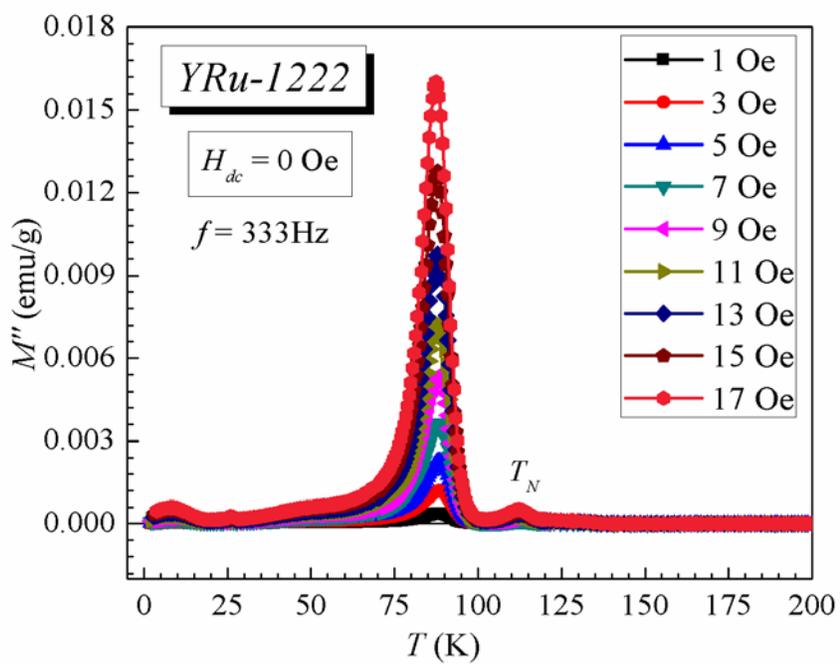

**Figure 9(a)**

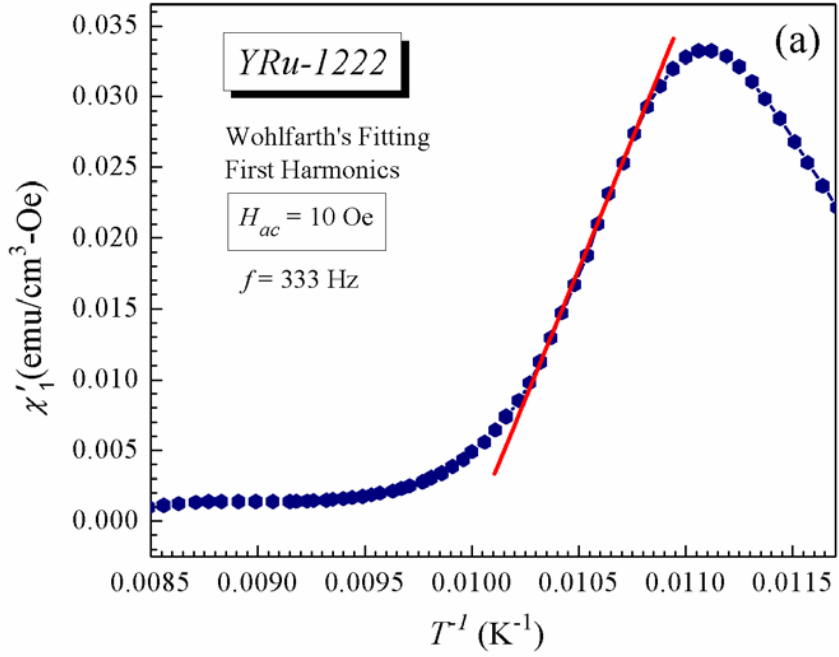

**Figure 9(b)**

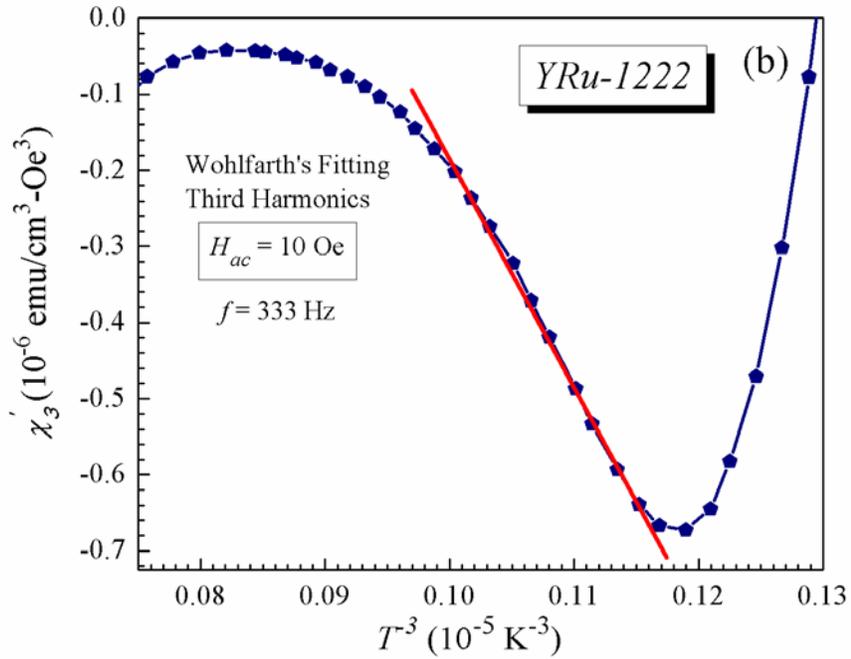

**Figure 10**

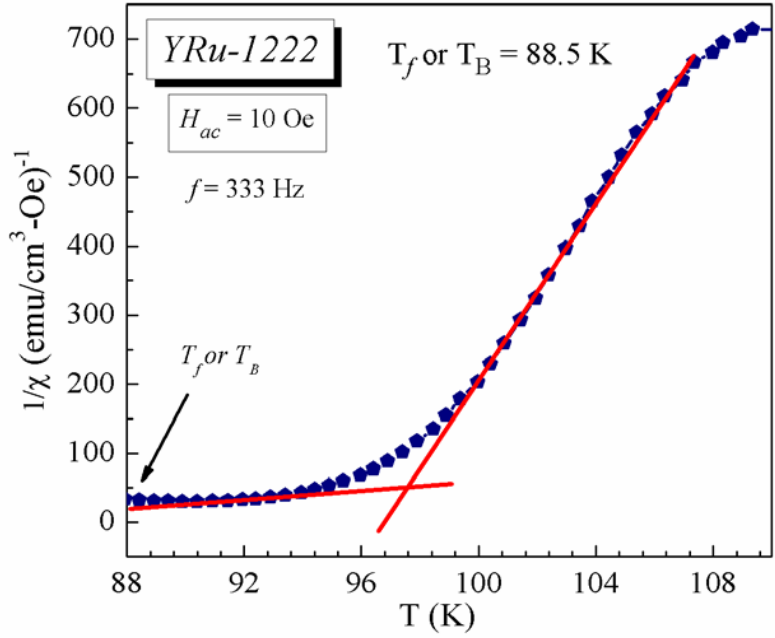

**Figure 11**

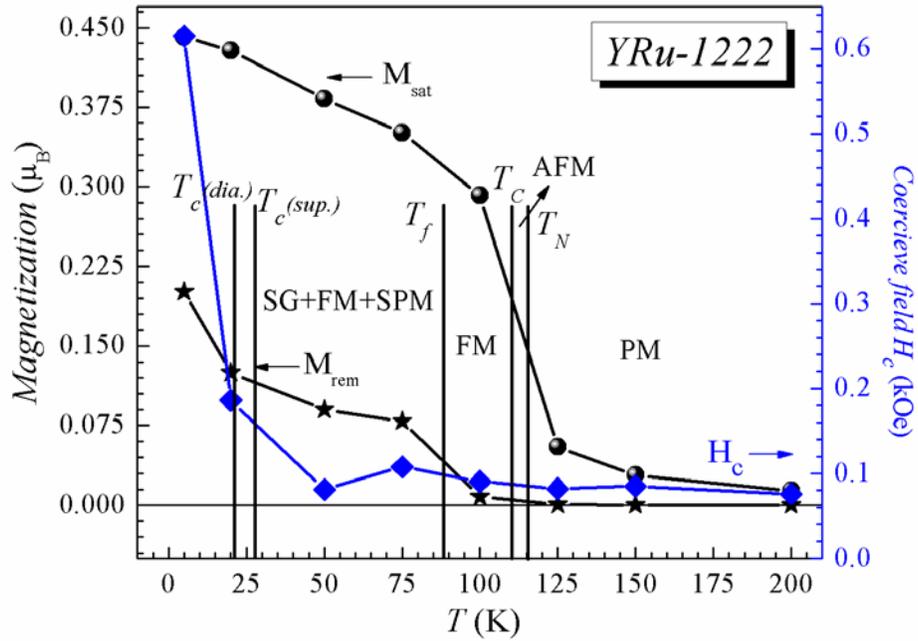